\documentclass[aps,prd,eprint,numerical,showpacs,showkeys,10pt]{revtex4-2}

\usepackage{longtable}
\usepackage{amsmath}
\usepackage{amssymb}
\usepackage{amsfonts}
\usepackage{graphicx}
\usepackage{dcolumn}
\usepackage{bm}
\usepackage{hyperref}
\usepackage{color}
\hypersetup{colorlinks=false}

\begin{document}

\title{Quasibound modes of rotating acoustic black holes}

\author{H. S. Vieira}
\email{horacio.santana.vieira@hotmail.com}
\email{horacio.santana.vieira@ifsc.usp.br}
\affiliation{Sao Carlos Institute of Physics, University of Sao Paulo, 13566-590, Sao Carlos, SP, Brazil}

\author{V. B. Bezerra}
\email{valdir@fisica.ufpb.br}
\affiliation{Departamento de F\'{i}sica, Universidade Federal da Para\'{i}ba, Caixa Postal 5008, 58051-970, Jo\~{a}o Pessoa, Paraíba, Brazil}

\date{\today}

\begin{abstract}
In this paper, we introduce a metric describing rotating acoustic black holes in spherical coordinates. This formulation allows us to obtain exact analytical solutions of the massless Klein--Gordon equation governing scalar perturbations in the corresponding background spacetime. By imposing the appropriate boundary conditions, we derive an exact analytical expression for the quasibound state frequencies of the system. Quasibound states play a central role in the dynamics of black hole analog systems, as they characterize long-lived resonant modes trapped in the vicinity of the effective horizon. Their investigation provides valuable insights into wave propagation, mode stability, and energy confinement in analog spacetimes. Moreover, because acoustic black holes offer experimentally accessible realizations of horizon physics, the study of quasibound states may establish a direct connection between theoretical predictions and laboratory observations, thereby contributing to a deeper understanding of phenomena that also arise in gravitational black hole environments.
\end{abstract}


\maketitle

%
%%%%%%%%%%%%%%%%%%%%%%%%%%%%%%%%%%%%%%%%% Introduction
%
\section{Introduction}
\label{Introduction}

In a seminal work, Unruh \cite{PhysRevLett.46.1351} demonstrated that, under suitable conditions, the propagation of linear acoustic perturbations in a fluid undergoing a transition from subsonic to supersonic flow can be described by a massless Klein--Gordon equation (KGE) in an effective Lorentzian spacetime. In this description, the background flow determines the effective acoustic metric, while the transition from subsonic to supersonic flow gives rise to a sonic horizon, providing an analog of the causal structure associated with a black hole horizon. This correspondence established the foundations of analog gravity and opened the possibility of investigating gravitational phenomena in laboratory systems through the dynamics of acoustic perturbations.

Since Unruh's original proposal, analog gravity has developed into a broad research field encompassing a variety of physical systems, including atomic condensates, flowing fluids, and other media in which perturbations propagate on an effective curved geometry. These systems have been used to investigate phenomena associated with horizons, such as Hawking radiation, superradiant amplification, and mode conversion, as well as the influence of dispersion on the propagation of perturbations near the horizon. Experimental observations of Hawking radiation in Bose--Einstein condensates (BECs) have provided an important realization of this programme, while theoretical studies have explored the extent to which horizon related phenomena persist beyond the idealized relativistic regime \cite{LivingRevRelativ.29.2,Philbin:2008kg,PhysRevD.79.064014,Steinhauer:2015saa,Steinhauer:2016gdn,Jacquet:2020feg}.

Although astrophysical and acoustic black holes are governed by fundamentally different dynamical equations, they can share important kinematical features. Astrophysical black holes arise as solutions of Einstein's field equations, whereas acoustic black holes emerge from the equations governing perturbations in a moving fluid. Nevertheless, both systems can possess horizons and support phenomena such as mode trapping, superradiant scattering, and Hawking radiation. These common features, together with the possibility of realizing acoustic horizons under controlled laboratory conditions, make analogue systems useful for investigating aspects of black hole physics that are difficult to access directly in astrophysical settings. Consequently, acoustic black holes have been studied in connection with Hawking radiation, wave scattering, superradiance, quasinormal modes, and other properties of effective curved spacetimes \cite{PhysRevD.44.1731,PhysRevD.48.728,PhysRevD.51.2827,ClassQuantumGrav.18.1137,GenRelGrav.34.1719,PhysRevA.78.021601,NewJPhys.10.103001,PhysRevD.81.124013,JHighEnergyPhys.04.030,PhysRevD.83.124016,PhysRevD.83.124047,JPhysBAtMolOptPhys.45.163001,PhysRevD.90.104015,PhysRevD.92.024043}.

Rotation introduces an additional level of richness into acoustic black hole geometries, since the azimuthal motion of the background flow modifies the effective spacetime and couples the perturbations to the angular momentum of the system. The rotating draining bathtub model (DBT) provided an early example of this class of analog geometries and has been extensively studied in connection with superradiance, quasinormal modes, and stability \cite{PhysRevD.59.124021,PhysRevD.66.024017,PhysRevD.70.124006,PhysRevD.70.124032}. More recently, rotating acoustic configurations have been investigated in more general settings, including slowly rotating curved geometries, where the spectra of quasinormal modes and quasibound states can be analyzed together with Hawking--Unruh radiation \cite{PhysRevD.105.045015}. Rotating acoustic metrics have also been considered in connection with the Sagnac effect and the geodesic structure of the effective spacetime \cite{EurPhysJC.84.565}.

Among the different spectral properties of acoustic black holes, quasibound states (QBSs) are of particular interest because they provide information about the effective potential and the confinement of perturbations outside the acoustic horizon. These modes are characterized by an ingoing boundary condition at the horizon and a decaying behavior at large distances, resulting in a discrete spectrum of complex frequencies. The study of QBSs therefore provides a complementary perspective to that offered by quasinormal modes and can reveal features of the effective geometry that are associated with the trapping of acoustic perturbations. In particular, the spectrum of QBSs has been obtained for Schwarzschild acoustic black holes and for analogue black holes realized in BECs \cite{PhysRevD.107.104038}.

The spectral analysis of rotating acoustic geometries is particularly relevant when the background flow contains both radial and azimuthal components. In such configurations, the circulation of the fluid determines the rotational properties of the effective geometry, while the radial flow controls the formation of the acoustic horizon. The resulting interplay between radial inflow and azimuthal motion affects the propagation of acoustic perturbations and, consequently, the associated QBS spectra. Recent studies of rotating acoustic geometries have shown that these spectra can be used to characterize the properties of the effective spacetime and to identify signatures associated with rotation \cite{PhysRevD.110.024035,PhysRevD.110.064084,PhysRevD.110.104058,PhysRevD.111.104064,EurPhysJC.85.570,PhysRevLett.135.243801,AnnPhys.494.170653}. This also motivates the investigation of more general rotating fluid configurations in which the idealized assumptions commonly adopted in acoustic black hole models can be relaxed.

In the present work, we investigate rotating acoustic black holes formulated in spherical coordinates. Although the DBT geometry is characterized by cylindrical symmetry, we show that the massless KGE can be formulated and solved analytically in spherical coordinates. In contrast to the standard cylindrical formulation, this representation allows the wave equation to be reduced to a form for which exact analytical solutions can be obtained. We then use these solutions to determine the spectrum of QBS frequencies and investigate the dependence of this spectrum on the parameters characterizing the acoustic geometry. Our results provide an analytical description of QBSs in the spherical representation of the DBT geometry and offer a complementary perspective on the spectral properties of rotating acoustic black holes (RABHs).

The paper is organized as follows. In Sec. \ref{RABH}, we derive the RABH metric in spherical coordinates following Unruh's approach. In Sec. \ref{KGE}, we obtain an analytical solution of the massless KGE in the resulting acoustic geometry. In Sec. \ref{QBS}, we impose the appropriate boundary conditions on the radial solution and determine the corresponding QBS spectrum. Finally, in Sec. \ref{Conclusions}, we summarize our main results and discuss their implications.

%
%%%%%%%%%%%%%%%%%%%%%%%%%%%%%%%%%%%%%%%%% Rotating acoustic black holes
%
\section{Rotating acoustic black holes}
\label{RABH}

In this section, we start by considering the general acoustic black hole solution in Minkowski spacetime obtained by Unruh~\cite{PhysRevLett.46.1351} and then we discuss a choice for the flow velocity in order to obtain RABHs in spherical coordinates. In fact, the well known metric for the DBT model is written in cylindrical coordinates \cite{ClassQuantumGrav.15.1767}, which could be called as laboratory coordinates. Indeed, we are going to (re)obtain an effective metric that is (more) similar to the one describing the Kerr spacetime background, which leads to an analytical study of field perturbations related to the QBSs.

The dynamics of an irrotational fluid is governed by the following fundamental equations of motions:
\begin{eqnarray}
\nabla \times \mathbf{v} & = & 0, \label{eq:irrotational}\\
\partial_{t}\rho+\nabla\cdot(\rho\mathbf{v}) & = & 0, \label{eq:Continuity}\\
\rho[\partial_{t}\mathbf{v}+(\mathbf{v}\cdot\nabla)\mathbf{v}] +\nabla p & = & 0. \label{eq:Euler}
\end{eqnarray}
In these equations, $\mathbf{v}$, $\rho$, and $p$ are the velocity, density, and pressure of the fluid, respectively. Next, we introduce the velocity potential $\Psi$, such that the velocity can be written as $\mathbf{v} = -\nabla \Psi$. In addition, we assume the fluid as barotropic, which implies that $\rho = \rho(p)$. Then, we proceed with the linearization of these equations of motion around some background configuration $(\rho_{0}, p_{0}, \Psi_{0})$, namely,
\begin{eqnarray}
\rho & = & \rho_{0}+\epsilon\rho_{1}, \label{eq:rho}\\
p & = & p_{0}+\epsilon p_{1}, \label{eq:p}\\
\Psi & = & \Psi_{0}+\epsilon\Psi_{1}, \label{eq:psi}
\label{eq2:Madelung_representation}
\end{eqnarray}
to get the following wave equation:
\begin{equation}
-\partial_{t}[c_s^{-2}\rho_{0}(\partial_{t}\Psi_{1}+\mathbf{v}_{0}\cdot\nabla\Psi_{1})]+\nabla\cdot[\rho_{0}\nabla\Psi_{1}-c_s^{-2} \rho_{0}\mathbf{v}_{0}(\partial_{t}\Psi_{1}+\mathbf{v}_{0}\cdot\nabla\Psi_{1})]=0.
\label{eq:wave_equation_Visser}
\end{equation}
Here, the local speed of sound $c_{s}$ is defined as
\begin{equation}
c_{s}^{-2} \equiv \frac{\partial \rho}{\partial p}.
\label{eq:sound}
\end{equation}
The wave equation~(\ref{eq:wave_equation_Visser}) describes the propagation of the phase fluctuations $\Psi_{1}$ as small excitations in a homogeneous and stationary condensate, which can be recast in the form of a wave equation in an effective curved spacetime as
\begin{equation}
\frac{1}{\sqrt{-g}}\partial_{\mu}(g^{\mu\nu}\sqrt{-g}\,\partial_{\nu}\Psi_{1})=0.
\label{eq:KG_equation}
\end{equation}
The above wave equation is similar to the covariant massless Klein--Gordon equation, where the associated acoustic line element can be written as
\begin{equation}
ds^{2} = g_{\mu\nu}dx^{\mu}dx^{\nu} = \frac{\rho_{0}}{c_{s}}[-c_{s}^{2}\,dt^{2}+(dx^{i}-v_{0}^{i}\,dt)\delta_{ij}(dx^{j}-v_{0}^{j}\,dt)].
\label{eq:acoustic_metric}
\end{equation}
Now, we assume the background flow as a spherically symmetric, stationary, and inviscid fluid with velocity $\mathbf{v}_{0}=v_{0}^{r}\mathbf{e}_{r}+v_{0}^{\phi}\mathbf{e}_{\phi}$, so that the acoustic metric given by Eq.~(\ref{eq:acoustic_metric}) becomes
\begin{equation}
\nonumber
ds^{2}=\frac{\rho_{0}}{c_{s}}\biggl\{-[c_{s}^{2}-(v_{0}^{r})^{2}-(v_{0}^{\phi})^{2}]\,dt^{2}+\frac{c_{s}}{c_{s}^{2}-(v_{0}^{r})^{2}}\,dr^{2}-2v_{0}^{\phi}r\sin\theta\,dt\,d\phi+r^{2}(d\theta^{2}+\sin^{2}\theta\,d\phi^{2})\biggr\},
\label{eq:spherical_acoustic_metric}
\end{equation}
where the following coordinate transformations were performed:
\begin{eqnarray}
&& dt \rightarrow dt-\frac{v_{0}^{r}}{c_{s}^{2}-(v_{0}^{r})^{2}}\,dr,\\
&& d\phi \rightarrow d\phi-\frac{v_{0}^{\phi}v_{0}^{r}}{r\sin\theta[c_{s}^{2}-(v_{0}^{r})^{2}]}\,dr.
\label{eq:coordinate_transformations}
\end{eqnarray}
The rotational nature of the acoustic geometry is reflected in the condition $c_{s}^{2}-(v_{0}^{r})^{2}-(v_{0}^{\phi})^{2}=0$, which determines the boundary of the ergoregion. The existence of such a region is a characteristic feature of rotating black hole spacetimes and is closely associated with mechanisms that allow the extraction of rotational energy, most notably superradiant scattering \cite{JETP.35.1085,NaturePhys.13.833,LectNotesPhys.971}. Its analogue in the present acoustic setting therefore raises the possibility of observing superradiant amplification of sound waves. Specifically, modes carrying suitable combinations of frequency and azimuthal angular momentum may gain energy as they interact with the rotating fluid, with the excess energy being supplied by the background flow. Consequently, the acoustic configuration considered here provides a natural platform for exploring superradiance in laboratory analogues of rotating black holes and may offer experimentally accessible means of probing this phenomenon under controlled conditions.

Next, the radial and angular components of the velocity field can be chosen as
\begin{eqnarray}
v_{0}^{r} & = & \frac{A}{r},\\\label{eq:radial_velocity}
v_{0}^{\phi} & = & \frac{J}{r} = \frac{C \sin\theta}{r},\label{eq:angular_velocity}
\end{eqnarray}
where $A$, $J$ and $C$ can be identified as the radial flow parameter, the angular circulation and the angular momentum, respectively. Therefore, these definitions allow us to write the acoustic line element for RABHs as
\begin{equation}
ds^{2} = -\biggl[f(r) - \frac{J^{2}}{r^{2}}\biggr]\,dt^{2} + \frac{1}{f(r)}\,dr^{2} - 2J\sin\theta\,dt\,d\phi + r^{2}(d\theta^{2} + \sin^{2}\theta\,d\phi^{2}),
\label{eq:metric_RABHs}
\end{equation}
where we have fixed $c_{s}=1$ and then dropped a position-independent prefactor by assuming a constant fluid density. The acoustic metric function $f(r)$ is given by
\begin{equation}
f(r) = 1-\frac{r_{h}^{2}}{r^{2}},
\label{eq:metric_function_RABHs}
\end{equation}
where we labeled the radial flow parameter as $A=r_{h}$, with $r_{h}$ being the horizon radius. Thus, $r_{e} = \sqrt{r_{h}^{2}+C^{2}}$ is the solution of the ergosphere equation $g_{tt}=0$, while $r_{\pm}=\pm r_{h}$ are the solutions of the surface equation $f(r)=0$. In fact, $r_{+}$ is the (exterior) acoustic event horizon, defined as the outermost marginally trapped surface for outgoing phonons at which the radial flow becomes supersonic. On the other hand, $r_{-}$ is a negative nonphysical solution. Furthermore, the effective gravitational acceleration at the acoustic (surface gravity) horizon is given by
\begin{equation}
\kappa_{h} \equiv \frac{1}{2}\frac{df(r)}{dr}\biggr|_{r = r_{h}} = \frac{1}{r_{h}},
\label{eq:surface_gravity_RABHs}
\end{equation}
where the last equality follows directly from the form of the metric function $f(r)$ given by Eq.~(\ref{eq:metric_function_RABHs}). In addition, the angular velocity associated with frame dragging in these RABHs is given by
\begin{equation}
\varpi(r) \equiv -\frac{g_{t\phi}}{g_{\phi\phi}} = \frac{C}{r^{2}},
\label{eq:dragging_RABHs}
\end{equation}
so that its value at the acoustic horizon is $\varpi_{h} \equiv \varpi(r_{h}) = C/r_{h}^{2}$, which characterizes the dragging of inertial frames at $r = r_{h}$.

The geometry defined by Eq.~\eqref{eq:metric_RABHs} possesses an acoustic horizon at $r=r_h$ for any choice of the parameters considered here, thereby providing an effective description of a rotating black hole analog. This feature is not necessarily realized in experimental vortex systems, where horizonless configurations may also occur. For instance, the experiments reported in Ref.~\cite{Nature.628.66} revealed two qualitatively different regimes, depending on the flow conditions: a configuration without a hollow core and a vortex state characterized by the formation of a hollow core. The latter presents a closer correspondence with black-hole-like geometries. Consequently, Eq.~\eqref{eq:metric_RABHs} may be regarded as a kinematical effective description of rotating vortex configurations in (super)fluids. While it reproduces the main geometric properties relevant to rotating acoustic spacetimes, it does not incorporate the nonlinear dispersive effects that arise in the experimental system of Ref.~\cite{Nature.628.66}.

We now turn to the propagation of acoustic perturbations outside the horizon and determine the associated quasibound state spectrum. Our analysis is based on the Vieira--Bezerra--Kokkotas (VBK) approach \cite{AnnPhys.373.28,PhysRevD.104.024035}, in which the relevant modes remain confined by an effective potential barrier and exhibit exponential decay in the asymptotic region. By introducing a suitable mode decomposition, the perturbation equation can be separated into angular and radial sectors, with the latter taking the form of a differential equation belonging to the Heun class.

%
%%%%%%%%%%%%%%%%%%%%%%%%%%%%%%%%%%%%%%%%% massless Klein--Gordon equation
%
\section{massless Klein--Gordon equation}
\label{KGE}

We proceed by examining the wave dynamics associated with the effective spacetime in Eq.~\eqref{eq:metric_RABHs}. Our primary objective is to characterize the acoustic perturbations in the exterior region and, subsequently, determine the corresponding QBS frequencies. The problem is governed by the massless KGE \eqref{eq:KG_equation}, which will be analyzed below in the background of the rotating acoustic geometry.

Despite its rotational character, the geometry described by Eq.~\eqref{eq:metric_RABHs} does not possess exact spherical symmetry, as can be seen from the explicit dependence of $g_{tt}$ on both $r$ and $\theta$. Nevertheless, its angular structure permits a decomposition analogous to that commonly employed in spherically symmetric configurations. We therefore consider the following representation for the scalar perturbation:
\begin{equation}
\Psi_{1}(t,r,\theta,\phi)\simeq e^{-i \omega t} R(r) P_{\nu}^{m}(\cos\theta) e^{i m \phi},
\label{eq:ansatz_RABHs}
\end{equation}
where $\omega$ is the mode frequency, $R(r)$ denotes the radial amplitude, and $P_{\nu}^{m}(\cos\theta)$ represents an associated Legendre function. Here, $m\in\mathbb{Z}_{+}$ is the azimuthal quantum number, whereas $\nu\in\mathbb{C}$ specifies the degree of the angular function. The corresponding separation constant is $\lambda=\nu(\nu+1)$. In the QBS problem, $\nu$ need not be restricted to integer values and may acquire a complex part, as will become relevant in the analysis below.

Using this decomposition in the massless KGE and evaluating the resulting expression for the metric given by Eq.~\eqref{eq:metric_RABHs}, the angular dependence can be isolated and the radial sector is governed by
\begin{equation}
R''(r)+\biggl[\frac{2}{r}+\frac{f'(r)}{f(r)}\biggr]R'(r)+\frac{r^{2}[\omega-m\varpi(r)]^{2}-\lambda f(r)}{r^{2}f^{2}(r)}R(r)=0,
\label{eq:radial_equation_RABHs}
\end{equation}
where primes denote differentiation with respect to $r$. Employing the explicit form of $f(r)$ and $\varpi(r)$ introduced previously, Eq.~\eqref{eq:radial_equation_RABHs} can be recast as
\begin{equation}
R''(r)+\biggl(\frac{1}{r-r_{h}}+\frac{1}{r+r_{h}}\biggr)R'(r)+\frac{(r^{2}\omega-Cm)^{2}-\lambda(r^{2}-r^{2}_{h})}{(r-r_{h})^{2}(r+r_{h})^{2}}R(r)=0,
\label{eq:radial_equation_RABHs_2}
\end{equation}
which will serve as the starting point for the subsequent QBS analysis.

The rotational contribution enters the radial dynamics through the combination $\omega-m\varpi_{h}$. Accordingly, the local threshold separating the superradiant and nonsuperradiant regimes is determined by the condition $\omega=\omega_{c}$, with
\begin{equation}
\omega_{c}=m\varpi_{h}.
\end{equation}
Thus, modes satisfying $\omega<\omega_{c}$ lie in the superradiant regime and can undergo amplification through rotational energy extraction.

%
%%%%%%%%%%%%%%%%%%%%%%%%%%%%%%%%%%%%%%%%% Quasibound state spectrum
%
\section{Quasibound state spectrum}
\label{QBS}

We begin by constructing an analytical solution to the radial equation \eqref{eq:radial_equation_RABHs_2} using the VBK approach. No boundary conditions are imposed at this stage, allowing us to obtain the general radial solution in terms of confluent Heun functions. A detailed account of this method can also be found in Refs.~\cite{EurPhysJC.82.669,EurPhysJC.82.932}. The physical QBS modes are subsequently selected by imposing an ingoing wave condition at the acoustic horizon and exponential decay at spatial infinity. These requirements define an eigenvalue problem for complex frequencies, whose solutions describe the oscillatory and damping properties of the QBS modes.

Following the VBK approach, Eq.~\eqref{eq:radial_equation_RABHs_2} can be transformed into the canonical confluent Heun equation,
\begin{equation}
y''(x)+\biggl(\alpha+\frac{1+\beta}{x}+\frac{1+\gamma}{x-1}\biggr)y'(x)+\biggl(\frac{\xi}{x}+\frac{\zeta}{x-1}\biggr)y(x)=0,
\label{eq:CHE_RABHs}
\end{equation}
where we introduce the dimensionless radial coordinate $x$, such that
\begin{equation}
x=\frac{r+r_{h}}{2r_{h}}
\label{eq:radial_coordinate_RABHs}
\end{equation}
and define the rescaled radial function $y(x)$ as
\begin{equation}
y(x)=x^{-\beta/2}(x-1)^{-\gamma/2}e^{-x\alpha/2}R(x).
\label{eq:radial_function_RABHs}
\end{equation}
The resulting solution is expressed in terms of the confluent Heun function,
\begin{equation}
y(x)=\mbox{HeunC}(\alpha,\beta,\gamma,\delta,\eta;x).
\label{eq:confluent_Heun_function_RABHs}
\end{equation}
The parameters characterizing this function depend on the mode frequency $\omega$, the azimuthal number $m$, the acoustic horizon radius $r_{h}$, and the circulation $C$ of the background flow. They are explicitly given by
\begin{eqnarray}
\alpha & = & -4ir_h\omega,\label{eq:alpha_RABHs}\\
\beta & = & -ir_h(\omega-\omega_c),\label{eq:beta_RABHs}\\
\gamma & = & -ir_h(\omega-\omega_c),\label{eq:gamma_RABHs}\\
\delta & = & 0,\label{eq:delta_RABHs}\\
\eta & = & \frac{r_h^2(\omega-\omega_c)(3\omega+\omega_c)}{2}-\lambda.
\label{eq:eta_RABHs}
\end{eqnarray}
Here, $\omega_c=m\varpi_h$ is the critical frequency evaluated at the acoustic horizon, with $\varpi_h=\varpi(r_h)$. The auxiliary coefficients $\xi$ and $\zeta$ entering Eq.~\eqref{eq:CHE_RABHs} are related to the Heun parameters through
\begin{eqnarray}
\xi & = & \frac{1}{2}(\alpha-\beta-\gamma+\alpha\beta-\beta\gamma)-\eta,\label{eq:xi_CHE}\\
\zeta & = & \frac{1}{2}(\alpha+\beta+\gamma+\alpha\gamma+\beta\gamma)+\delta+\eta.
\label{eq:zeta_CHE}
\end{eqnarray}
The general solution of the radial equation can be expressed as a linear combination of two linearly independent confluent Heun solutions.

To identify the QBS modes, we examine the asymptotic behavior of the radial solutions near the acoustic event horizon and in the far-field region. The physically relevant modes must satisfy a purely ingoing wave condition at the horizon and decay exponentially at spatial infinity. Within the VBK approach, these requirements are associated with polynomial truncation conditions for the confluent Heun function. Such truncation requires the simultaneous fulfillment of two constraints,
\begin{eqnarray}
\frac{\delta}{\alpha}+\frac{\beta+\gamma+2}{2}+n & = & 0,\label{eq:delta-condition}\\
\Delta_{n+1}(\xi) & = & 0,\label{eq:Delta-condition}
\end{eqnarray}
where $n=0,1,2,\ldots$ is the polynomial degree, and $\Delta_{n+1}(\xi)$ denotes the determinant condition associated with the confluent Heun equation. The resulting quantization conditions provide the complete set of allowed complex frequencies characterizing the QBSs. For the parameters obtained above, the first constraint, commonly referred to as the $\delta$-condition, yields a quantization relation for the mode frequency. Substituting $\delta=0$ and $\beta=\gamma=-ir_h(\omega-\omega_c)$ into Eq.~\eqref{eq:delta-condition}, we obtain
\begin{equation}
\omega_{mn}=\omega_c-\frac{i(n+1)}{r_{h}},
\label{eq:omega_RABHs}
\end{equation}
where $n(=0,1,2,\ldots)$ is the overtone number. The frequency relation in Eq.~\eqref{eq:omega_RABHs} exhibits the symmetry
\begin{equation}
\omega_{mn}=-[\omega_{-mn}]^*,
\label{eq:symmetry_RABHs}
\end{equation}
where the asterisk denotes complex conjugation. This relation follows from the linear dependence of $\omega_c$ on $m$, provided the remaining background parameters are held fixed. It implies that modes with opposite azimuthal numbers have identical damping rates, whereas their oscillation frequencies have opposite signs. Thus, the corotating and counterrotating branches are related by reflection across the imaginary axis in the complex-frequency plane. This symmetry holds for arbitrary values of the circulation parameter $C$ and acoustic horizon radius $r_h$, within the parameter domain of the model.

An additional consequence of Eq.~\eqref{eq:omega_RABHs} is that the real part of the candidate QBS frequencies coincides with the critical frequency at the horizon,
\begin{equation*}
\mbox{Re}(\omega_{mn})=\omega_c=m\varpi_h.
\end{equation*}
The corresponding modes are therefore synchronized with the acoustic horizon. In rotating geometries, the critical frequency marks the threshold separating ordinary absorption from superradiant amplification. At synchronization, the energy flux associated with the rotational interaction at the horizon vanishes, marking the boundary between these two regimes. This condition should not, however, be interpreted as the absence of all possible wave--background interactions.

\begin{figure}[t]
\centering
\includegraphics[scale=1]{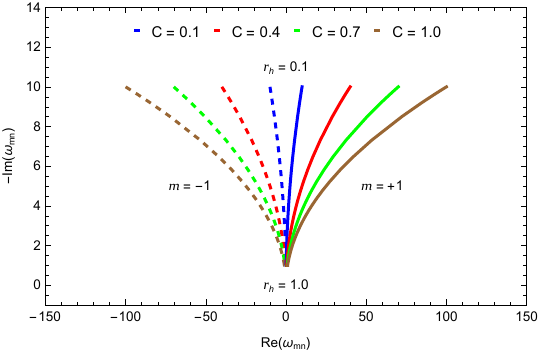}
\caption{Fundamental ($n=0$) QBS frequencies for different acoustic horizon radii $r_h$ and circulation parameters $C$. Solid and dashed curves represent corotating ($m>0$) and counterrotating ($m<0$) modes, respectively.}
\label{fig:Fig1_QBSs_RABHs}
\end{figure}

Figure~\ref{fig:Fig1_QBSs_RABHs} displays the fundamental ($n=0$) QBS frequencies. The symmetry discussed above relates the corotating ($m>0$) and counterrotating ($m<0$) branches: their imaginary components coincide, while their real components have opposite signs. Consequently, the two branches are mirror images under the transformation $\operatorname{Re}(\omega_{m0})\rightarrow-\operatorname{Re}(\omega_{-m0})$. This correspondence is preserved for different circulation parameters $C$ and acoustic horizon radius $r_h$, provided the same spectral conditions are satisfied.

\begin{figure}[t]
\centering
\includegraphics[scale=1]{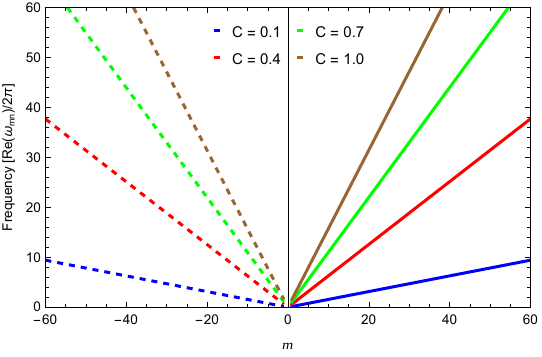}
\caption{Two-dimensional frequency spectra of the fundamental ($n=0$) QBS frequencies for a unit acoustic horizon radius, $r_h=1$, and different circulation parameters $C$. Solid and dashed curves represent corotating ($m>0$) and counterrotating ($m<0$) modes, respectively.}
\label{fig:Fig2_QBSs_RABHs}
\end{figure}

Figure~\ref{fig:Fig2_QBSs_RABHs} presents the predicted frequency distributions of the fundamental ($n=0$) QBS modes for different circulation parameters. These results may be considered alongside the experimental observations reported in Fig.~2(f,g) of Ref.~\cite{Nature.628.66}. The plotted spectra exhibit a low-frequency region without QBS excitations, with the first green curve marking the lower-frequency boundary of the displayed modes. This feature suggests the presence of a frequency threshold within the parameter range considered. However, interpreting this boundary as a general propagation threshold requires a separate analysis of the admissible modes and their boundary conditions.

\begin{figure}[t]
\centering
\includegraphics[scale=1]{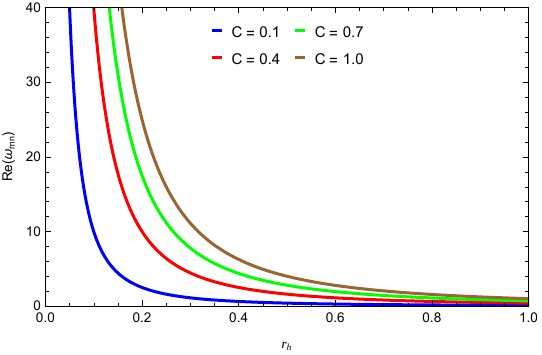}
\caption{Real part of the fundamental ($n=0$) QBS frequencies as a function of the acoustic horizon radius $r_h$, for different circulation parameters $C$ and corotating modes with $m=1$.}
\label{fig:Fig3_QBSs_RABHs}
\end{figure}

The dependence of the real part of the fundamental QBS frequencies on the acoustic horizon radius is illustrated in Fig.~\ref{fig:Fig3_QBSs_RABHs}. For fixed circulation $C$ and corotating modes with $m=1$, the oscillation frequency follows the critical frequency, $\omega_c=C/r_h^2$. Hence, the analytical expression predicts an inverse-square dependence on the horizon radius, with the oscillation frequency decreasing as $r_h$ increases for positive $C$. Similar structured frequency trends have been reported in experiments involving a giant hollow-core quantum vortex in superfluid $^4$He~\cite{Nature.628.66}. Despite this qualitative resemblance, a direct quantitative comparison is not currently justified, since the experimentally studied vortex lacks radial flow and therefore does not possess an acoustic horizon. The present model describes a distinct configuration in which radial draining and circulation coexist, providing a framework for investigating how horizon-related effects influence the spectrum of acoustic perturbations.

\begin{figure}[t]
\centering
\includegraphics[scale=1]{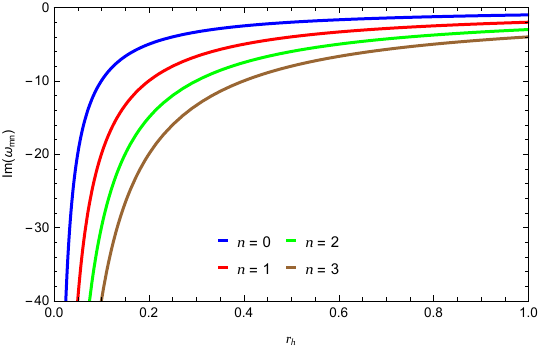}
\caption{Damping rates of the QBS frequencies as functions of the acoustic horizon radius $r_h$, for different overtone numbers $n$ and corotating modes with $m=1$.}
\label{fig:Fig4_QBSs_RABHs}
\end{figure}

Finally, Fig.~\ref{fig:Fig4_QBSs_RABHs} shows the imaginary parts of the QBS frequencies as functions of the acoustic horizon radius. According to Eq.~\eqref{eq:omega_RABHs}, the imaginary component is given by $\operatorname{Im}(\omega_{mn})=-(n+1)/r_h$. Thus, the damping rate depends exclusively on the horizon radius and the overtone number, remaining independent of the circulation parameter. The resulting inverse dependence on $r_h$ accounts for the hyperbolic behavior displayed in the figure.

Furthermore, the values for the separation constant $\lambda$, as well as for the general degree $\nu$, which satisfies the second constraint given by Eq.~(\ref{eq:Delta-condition}), commonly referred to as the $\Delta$-condition, are shown in Table \ref{tab:I_RABHs}.

\begin{table}[t]
\caption{Fundamental QBSs frequencies $\omega_{mn}$ for different acoustic radii $r_{h}$ and circulation parameters $C$, and their corresponding separation constants $\lambda_{mn;s}$ and general degrees $\nu_{mn;s}$, with $1 \leq s \leq n+1$.}
	\label{tab:I_RABHs}
	\begin{tabular}{c|c|c|c|c}
		\hline\noalign{\smallskip}
		$r_{h}$ & $C$   & $\omega_{m0}$ & $\lambda_{m0;1}$ & $\nu_{m0;1}$			\\
		\noalign{\smallskip}\hline\noalign{\smallskip}
		\multicolumn{5}{c}{$m=1$} 			\\
		\noalign{\smallskip}\hline\noalign{\smallskip}
		$0.1$   & $1.0$ & $100-10i$     & $-2-20i$        & $2.52708-3.30351i$ \\
		$0.4$   & $1.0$ & $6.25-2.5i$   & $-2-5i$         & $0.83180-1.87715i$ \\
		$0.7$   & $1.0$ & $2.04082-1.42857i$ & $-2-2.85714i$ & $0.39456-1.59695i$ \\
		$1.0$   & $1.0$ & $1-i$ & $-2-2i$ & $0.17362-1.48451i$ \\
		\noalign{\smallskip}\hline\noalign{\smallskip}
		\multicolumn{5}{c}{$m=2$} 			\\
		\noalign{\smallskip}\hline\noalign{\smallskip}
		$0.1$   & $1.0$ & $200-10i$     & $-2-40i$        & $3.87540-4,57101i$ \\
		$0.4$   & $1.0$ & $12.5-2.5i$   & $-2-10i$        & $1.54963-2.43946i$ \\
		$0.7$   & $1.0$ & $4.08163-1.42857i$ & $-2-5.71429i$ & $0.95365-1.96548i$ \\
		$1.0$   & $1.0$ & $2-i$ & $-2-4i$ & $0.64369-1.74872i$ \\
		\noalign{\smallskip}\hline
	\end{tabular}
\end{table}

%
%%%%%%%%%%%%%%%%%%%%%%%%%%%%%%%%%%%%%%%%% Final remarks
%
\section{Final remarks}
\label{Conclusions}

Analog gravity provides a versatile framework for studying gravitational phenomena through experimentally accessible systems whose effective dynamics reproduce characteristic features of curved spacetimes. In this work, we developed an effective rotating acoustic geometry and investigated the dynamics of acoustic perturbations in its exterior region. By solving the corresponding massless Klein--Gordon equation analytically, we obtained the quasibound-state frequencies associated with the rotating acoustic black hole. The resulting spectrum contains both corotating and counterrotating modes and exhibits features that are qualitatively related to those observed in previous experiments involving hollow-core quantum vortices and other analog-gravity systems \cite{Nature.628.66,PhysRevLett.121.061101,PhysRevLett.125.011301}.

The QBS analysis provides a direct characterization of the mode structure supported by the effective geometry. In particular, the spectrum contains modes with opposite azimuthal quantum numbers, allowing the corotating and counterrotating sectors to be considered within the same framework. The corresponding frequencies exhibit the symmetry $\omega_{mn}=-[\omega_{-mn}]^*$, while their damping rates are controlled by the acoustic horizon radius and the overtone number. Moreover, the real part of the QBS frequency coincides with the critical frequency associated with the rotation of the acoustic horizon, indicating that the quasibound modes are synchronized with the horizon.

The framework developed here also suggests several extensions of the present analysis. One possibility is to incorporate environmental effects into rotating acoustic configurations, in analogy with the role played by astrophysical environments in black-hole perturbation theory. For instance, an experimental realization based on a hollow-core vortex with nonzero circulation and radial flow could provide a setting in which an acoustic horizon is present. Additional modifications of the background flow could then be introduced through external impurities, provided that the resulting perturbations do not destroy the vortex configuration, or through nonlinear dispersive interactions in a superfluid. Such extensions could be used to investigate effective geometries containing additional structures analogous, at the level of the perturbation dynamics, to accretion environments, black-hole hair \cite{PhysRevD.103.044042,GenRelGrav.54.49}, or dark-matter distributions surrounding compact objects \cite{PhysRevD.105.L061501,PhysRevD.107.084027}.

More realistic analog configurations may therefore provide a controlled setting in which the influence of environmental and dissipative effects on acoustic spectra can be investigated. Such studies could contribute to the broader understanding of wave propagation in strong-field analog systems and to comparisons between theoretical predictions and laboratory measurements \cite{PhysRevD.111.024017,PhysRevD.111.064026}. In this context, the spatial dependence of the radial eigensolutions obtained here may also be useful for designing future experiments aimed at probing the response of rotating acoustic systems in the presence of dissipation.

Finally, the relation $\operatorname{Re}(\omega_{mn})=\omega_c$ obtained for the QBS frequencies has a direct interpretation in terms of rotational synchronization. The critical frequency defines the threshold associated with the exchange of energy between a wave and a rotating horizon. At this frequency, the wave is synchronized with the angular motion of the acoustic horizon, and the horizon contribution to the energy flux associated with rotational amplification vanishes. Thus, the QBS spectrum obtained here naturally lies at the synchronization threshold of the rotating acoustic geometry, providing a direct connection between the analytic spectrum and the superradiant properties of the background.

%
%%%%%%%%%%%%%%%%%%%%%%%%%%%%%%%%%%%%%%%%% Acknowledgments
%
\begin{acknowledgments}

This study was financed in part by the Conselho Nacional de Desenvolvimento Científico e Tecnológico -- CNPq -- Research Project No. 446211/2024-9 and Research Fellowship No. 314926/2025-9. V. B. B. is partially supported by the CNPq -- Grant No. 311847/2026-9.

\end{acknowledgments}
%
%%%%%%%%%%%%%%%%%%%%%%%%%%%%%%%%%%%%%%%%% DATA AVAILABILITY
%
\section*{DATA AVAILABILITY}

There are no publicly available research data or software supporting this manuscript. Requests for further information or data should be sent to the authors.

%
%%%%%%%%%%%%%%%%%%%%%%%%%%%%%%%%%%%%%%%%% References
%

%
%%%%%%%%%%%%%%%%%%%%%%%%%%%%%%%%%%%%%%%%%%%%%%%%%%%%%%%%%%%%%%%%%%%%%%%%%%%%%%%%%%%%%%%%%%%%%%
%
\end{document}